**The Australian Aboriginal People: How to Misunderstand Their Science**
Ray Norris, CSIRO Astronomy & Space Science, Sydney, Australia

The following popular article was published in "The Conversation" in April 2014:
[http://theconversation.com/aboriginal-people-how-to-misunderstand-their-science-23835]

**Introduction**

Just one generation ago, schoolkids were taught that Aboriginal people couldn't count beyond five, wandered the desert scavenging for food, had no civilization or religion, had no agriculture, couldn't navigate, didn't build houses, and peacefully acquiesced when Western Civilisation rescued them in 1788.

**How did we get it so wrong?**

Bill Gammage (1) and others have shown that land was carefully managed to maximize productivity, resulting in fantastically fertile soils, now exploited and almost destroyed by intensive agriculture. In some cases, people built villages with stone houses, and planted crops. They had sophisticated number systems(2), knew bush medicine, knew the cardinal points to a high accuracy, could navigate using stars and oral maps (3) and used these navigational skills to support flourishing trade routes across the country. They mounted fierce resistance to the British invaders, in some cases winning significant military victories (4). Funny how these stories of the Aboriginal resistance against the invaders were not taught in Australian schools until the last decade or so.

Only now are we starting to understand Aboriginal intellectual and scientific achievements. The Yolngu people, in north eastern Arnhem Land in the Northern Territory, long recognised (5) how the tides are linked to the phases of the moon. Back in the early 17th century, Italian scientist Galileo Galilei was still proclaiming, incorrectly, that the moon had nothing to do with tides (6).

In some cases, Aboriginal people had figured out how eclipses work, and were well aware of the how the planets moved differently from the stars. They used this knowledge (7) for navigation, to construct calendars, and to regulate the cycles of travel from one place to another to maximize the availability of seasonal foods.

**Why are we only finding this out now?**

We owe much of our knowledge about pre-contact Aboriginal culture to the great anthropologists of the 20th century. Their massive tomes tell us much about Aboriginal art, songs, and spirituality but are strangely silent about intellectual achievements, such as Aboriginal understanding of how the world works, or how they navigated. For example, in his 1938 book "*The Australian Aborigines: How to Understand Them*" Elkin appears to have heard at least one songline (an oral map) without noting its significance:

*"[...] its cycle of the hero's experiences as he journeyed from the north coast south and then back again north... a headman sitting nearby commented that the Ngurlmak, according to the text, was now in that country, then in another place, and so on, ever coming nearer until at last it was just where we were making the recording".*

Mountford also encountered descriptions of songlines, but didn't remark on their navigational significance. How could these giants of anthropology not recognise the significance of what they had been told?

The answer dawned on me recently, when I gave a talk (8) on Aboriginal Navigation at the National Library of Australia, and posed this question to the audience.

Afterwards a lady who had been one of Elkin's PhD students told me that Elkin worked within fixed ideas about what constituted Aboriginal culture. I realised that she was describing exactly what the American philosopher of science Thomas Kuhn [9] was referring to when he coined the name "paradigm".

**The Paradigm Problem**

According to Kuhn, all of us, including scientists and anthropologists, are fallible humans. We grow up with a paradigm (e.g. Aboriginal culture is "primitive") which we accept as true, and anything that doesn't fit into that paradigm is dismissed as irrelevant or aberrant.

Only 200 years ago, people discussed whether Aboriginal people were "sub-human"(10). Ideas change slowly, over generations, and the underlying message lingers on long after it has been demonstrably falsified. As late as 1923, W.J. Thomas (11) could write "When the white man first carried the burden and blessing of civilisation to the shores of Australia, he found the land inhabited by a very primitive race of people". And even in the 1960's Rolf Harris (12) could sing with impunity: " Let me Abo's go loose, They're of no further use"

**Not so Primitive**

The prevailing paradigm in Elkin's time was that Aboriginal culture was primitive, and inferior to European cultures, and so couldn't possibly have anything useful to say about how the world worked, or how to manage the land, or how to navigate.

So an anthropologist might study the Aboriginal people as objects, or anthropological curiosities, just as a biologist might study insects under a microscope, but would have nothing to learn from Aboriginal people themselves. Even now, the paradigm lives on. Well-educated white Australians, even those who try so hard to be politically correct, find it very hard to escape their childhood image of "primitive" Aboriginal people.

We must overcome the intellectual inertia that keeps us in that old paradigm, preventing us from recognising the enormous contribution that Aboriginal culture can make to our own understanding of the world, and to our attempts to manage it. As Thomas Kuhn said (13) *"…when paradigms change, the world itself changes with them."*

**Still to learn**

In recent years, it has become clear that traditional Aboriginal people knew a great deal (14) about the sky, knew the cycles of movements of the stars and the complex motions of the sun, moon and planets.

There is even found a sort of "Aboriginal Stonehenge"(15), that points to the sunset on midsummers day and midwinters day. And I suspect that this is only the tip of the iceberg of Aboriginal astronomy.

So in the debate about whether our schools should include Aboriginal perspectives (16) in their lessons, I argue that kids studying science today could also learn much from the way that pre-contact Aboriginal people used observation to build a self-consistent picture of the world around them, with predictive power and practical applications.

This "ethno-science" is very similar to modern science in many respects, but is of course couched in appropriate cultural terms and without the iconic billion-dollar telescopes and particle accelerators.

But if you want to learn about the essence of how science works, how people build a self-consistent picture of the world to solve practical problems, the answer may be clearer in an Aboriginal community than in a high-tech laboratory.